\documentstyle[aps,preprint]{revtex}

\begin{document}
\draft
\title{A new topological aspect of the arbitrary dimensional topological defects}
\author{Ying Jiang\thanks{E-mail: yjiang@itp.ac.cn}}
\address{CCAST (World Laboratory), Box 8730, Beijing 100080, P. R. China}
\address{Institute of Theoretical Physics, Chinese Academy of Sciences, P. O. Box
2735,\\
Beijing 100080, P.R. China\thanks{%
mailing address}
}
\author{Yishi Duan\thanks{%
E-mail: ysduan@lzu.edu.cn}}
\address{Institute of Theoretical Physics, Lanzhou University, Lanzhou
730000, P. R. China}
\maketitle

\begin{abstract}
We present a new generalized topological current in terms of the order
parameter field $\vec \phi$ to describe the arbitrary
dimensional topological defects. By virtue of the $%
\phi$-mapping method, we show that the topological defects are generated
from the zero points of the order parameter field $\vec \phi$, and the
topological charges of these
topological defects are topological quantized in terms of the Hopf indices
and Brouwer degrees of $\phi$-mapping under the condition that the Jacobian $%
J(\frac \phi v)\neq 0$. When $J(\frac \phi v)=0$, it is shown that there
exist the crucial case of branch process. Based on the implicit function
theorem and the Taylor expansion, we detail the bifurcation of generalized
topological current and find different directions of the bifurcation. The
arbitrary dimensional topological defects are found splitting or merging at
the degenerate point of field function $\vec \phi $ but the total charge of
the topological defects is still unchanged.
\end{abstract}




\section{Introduction}

The world of topological defects is amazingly rich and have been the focus
of much attention in many areas of contemporary physics\cite
{hindmarsh1,vilenkin1,gueron1}. The importance of the role of defects in
understanding a variety of problems in physics is clear\cite
{deboer1,su1,blanco1,cen1}. Whenever we have a field theory with a set of
vacua given by a non-connected space there is the possibility of having
different regions in space living on different vacuum sectors. Two such
regions will meet at what is generally named a topological defect, i.e. a
thin hypersurface where the field rapidly evolves from one vacuum to the
other\cite{paris1}.

As a edivence of cosmological phase transitions in the early Universe,
topological defects, such as cosmic string, domain wall, monopole and
texture, remain somewhere in our Universe, and can help to resolve some long
standing puzzles such as the origin of structure formation\cite{delaix1}.
The physics of spacetimes containing the defects has been investigated
extensively\cite{vilenkin2}. The existence of topological defects\cite
{vilenkin1,vilenkin2} with a non-trivial core phase structure has recently
been demonstrated for embeded global domain walls and vortices\cite
{axenides1}. A solution for a Schwarzschild particle with global monopole
charge has been obtained by Barriola and Vilenkin\cite{vilenkin1}. The
texture model of structure formation in the universe\cite{turok2} and the
one--texture universe\cite{davis3} have been studied by many researchers\cite
{chen1}. $p$--branes\cite{townsend1}, which have been found to play
important roles in $M$--theory\cite{dai1}, are also proved to be topological
defects in gauge theory\cite{diamantini1}. Recently, some physicists noticed%
\cite{lalak1} that the topological defects are closely related to the
spontaneously broken of $O(m)$ symmetry group to $O(m-1)$ by $m$--component
order parameter field $\vec{\phi}$ and pointed out that for $m=1$, one has
domain walls, $m=2$, strings and $m=3$, monopoles, for $m=4$, there are
textures. And $O(m)$ symmetric vector field theories are a class of models
describing the critical behavior of an great variety of important physical
systems\cite{nuno,halperin}. But for the lack of a powerful method, the topological
properties of these systems are not very clear, some important topological
informations have been lost, also the unified theory of describing the
topological properties of all these defect objects is not established yet.

In this paper, in the light of $\phi $--mapping topological current theory%
\cite{duan3}, a useful method which plays a important role in studying the
topological invariants\cite{duan4,duan8} and the topological structures of
physical systems\cite{duan6,duan9}, we will investigate the topological
quantization and the branch process of arbitrary dimensional topological
defects. We will show that the topological defects are generated from the
zero points of the order parameter field $\vec \phi$, and their topological
charges are quantized in terms of the Hopf indices and Brouwer degrees of $%
\phi$-mapping under the condition that the zero points of field $\vec \phi$
are regular points. While at the critical points of the order parameter
field $\vec{\phi}$, i.e. the limit points and bifurcation points, there
exist branch processes, the topological current of defect bifurcates and the
topological defects split or merge at such point, this means that the
topological defects system is unstable at these points.

This paper is organized as follows. In section 2, we investigate the
topological quantization of these topological defect and point out that the
topological charges of these defects are the Winding numbers which are
determined by the Hopf indices and the Brouwer degrees of the $\phi $%
--mapping. In section 3, we study the branch process of the defect
topological current at the limit points, bifurcation points and higher
degenerated points systematically by virtue of the $\phi $--mapping theory
and the implicit function theorem.

\section{Topological quantization of topological defects}

In our previous papers\cite{duan3,duan9,jiang1,jiang2,jiang3}, we have
studied the topological properties of point like defects and string like
defects systematically via the $\phi$-mapping topological current theory and
rank-2 topological current theory, respectively. In this paper, in order to
study the topological properties of arbitrary dimensional topological
defects, we will extend the concept to present an arbitrary dimensional
generalized topological current. 

It is well known that the $m$--component vector order parameter field 
$\vec \phi (x)=(\phi ^1 (x),...,\phi ^m (x))$
determines the properties of the topological defect system, and it can be looked
upon as a smooth mapping between the $n$--dimensional Riemannian spacetime $G$ 
(with the metric tensor $g_{\mu \nu }$ and local coordinates $x^\mu $ ($\mu ,\nu =1,...,n$)) 
and a $m$--dimensional Euclidean space $R^m$ as $\phi :\;G\rightarrow R^m$.
By analogy with the discussion in our previous work\cite{duan9,jiang1,jiang2,jiang3}, from this 
$\phi$-mapping, one can deduce a topological tensor current as
\begin{equation}  \label{4.8}
j^{\mu _1\cdot \cdot \cdot \mu _k}=\frac 1{A(S^{m-1})(m-1)!\sqrt{g_x}}%
\epsilon ^{\mu _1\cdot \cdot \cdot \mu _k\mu _{k+1}\cdot \cdot \cdot \mu
_n}\epsilon _{a_1\cdot \cdot \cdot a_m}\partial _{\mu _{k+1}}n^{a_1}\partial
_{\mu _{k+2}}n^{a_2}\cdot \cdot \cdot \partial _{\mu _n}n^{a_m}.
\end{equation}
to describe the system of the topological defects, where $k=n-m$. In this expression, 
$\partial_\mu$ stands for $\partial/ \partial x^\mu$, 
$A(S^{m-1})=2\pi ^{m/2}/\Gamma (m/2)$
is the area of $(m-1)$--dimensional unit sphere $S^{m-1}$ and $n^a (x)$ is 
the direction field of the $m$--component order parameter field $\vec \phi$
\begin{equation}
n^a(x)=\frac{\phi ^a(x)}{||\phi (x)||},\;\;\;\;\;||\phi (x)||=\sqrt{\phi
^a(x)\phi ^a(x)}  \label{4.0}
\end{equation}
with $n^a(x)n^a(x)=1$. It is obviously that $n^a(x)$ is a section of the 
sphere bundle $S(G)$\cite{duan3} and it can be looked upon as a map of $G$
onto a $(m-1)$--dimensional unit sphere $S^{m-1}$ in order parameter space.
Clearly, the zero points of the order parameter field $\vec{\phi}(x)$
are
just the singular points of the unit vector $n^a(x)$. 
It is easy to see that $j^{\mu _1\cdot \cdot \cdot \mu _k}$ are completely
antisymmetric, and from the formulas above, we conclude that there exists a
conservative equation of the topological tensor current in (\ref{4.8})
\[
\nabla _{\mu _i}j^{\mu _1\cdot \cdot \cdot \mu _k}=0,\;\;\;i=1,...,k. 
\]

In the following, we will investigate the intrinsic structure
of the generalized topological current $j^{\mu _1\cdot \cdot \cdot
\mu _k}$ by making use of the $\phi $--mapping method. From (\ref{4.0}), we have 
\[
\partial _\mu n^a=\frac 1{||\phi ||}\partial _\mu \phi ^a+\phi ^a\partial
_\mu (\frac 1{||\phi ||}),\;\;\;\frac \partial {\partial \phi ^a}(\frac 1{%
||\phi ||})=-\frac{\phi ^a}{||\phi ||^3} 
\]
which should be looked upon as generalized functions\cite{gelfand1}. Due to
these expressions the generalized topological current (\ref{4.8}) can be
rewritten as 
\begin{eqnarray}
j^{\mu _1\cdot \cdot \cdot \mu _k} &=&C_m\frac 1{\sqrt{g_x}}\epsilon ^{\mu
_1\cdot \cdot \cdot \mu _k\mu _{k+1}\cdot \cdot \cdot \mu _n}\epsilon
_{a_1\cdot \cdot \cdot a_m}  \nonumber \\
&\cdot& \partial _{\mu _{k+1}}\phi ^a\cdots \partial _{\mu _n}\phi ^{a_m}%
\frac \partial {\partial \phi ^a}\frac \partial {\partial \phi ^{a_1}}%
(G_m(||\phi ||)),
\end{eqnarray}
where $C_m$ is a constant 
\[
C_m=\left\{ 
\begin{array}{cc}
-\frac 1{A(S^{m-1})(m-2)(m-1)!}, & \;\;\;\;m>2 \\ 
\frac 1{2\pi }, & \;\;\;\;m=2
\end{array}
,\right. 
\]
and $G_m(||\phi ||)$ is a Green function 
\[
G_m(||\phi ||)=\left\{ 
\begin{array}{ccc}
\frac 1{||\phi ||^{m-2}} & ,\;\;\; & m>2 \\ 
\ln ||\phi || & ,\;\;\; & m=2
\end{array}
.\right. 
\]
Defining general Jacobians $J^{\mu _1\cdot \cdot \cdot \mu _k}(\frac \phi x)$
as following 
\[
\epsilon ^{a_1\cdot \cdot \cdot a_m}J^{\mu _1\cdot \cdot \cdot \mu _k}(\frac %
\phi x)=\epsilon ^{\mu _1\cdot \cdot \cdot \mu _k\mu _{k+1}\cdot \cdot \cdot
\mu _n}\partial _{\mu _{k+1}}\phi ^{a_1}\partial _{\mu _{k+2}}\phi
^{a_2}\cdot \cdot \cdot \partial _{\mu _n}\phi ^{a_m} 
\]
and by making use of the $m$--dimensional Laplacian Green function relation 
in $\phi$--space\cite{duan3} 
\[
\Delta _\phi (G_m(||\phi ||))=-\frac{4\pi ^{m/2}}{\Gamma (\frac m2%
-1)}\delta (\vec{\phi}) 
\]
where $\Delta _\phi =(\frac{\partial ^2}{\partial \phi ^a\partial \phi ^a})$
is the $m$--dimensional Laplacian operator in $\phi $--space, we do obtain
the $\delta $--function structure of the defect topological current rigorously 
\begin{equation}
j^{\mu _1\cdot \cdot \cdot \mu _k}=\frac 1{\sqrt{g_x}}\delta (\vec{\phi}%
)J^{\mu _1\cdot \cdot \cdot \mu _k}(\frac \phi x).  \label{4.10}
\end{equation}
This expression involves the total defect information of the system and 
it indicates that all the defects are located at the zero points of the order
parameter field $\vec \phi (x)$. It must be pointed out that, comparing to
similar expressions in other papers, the results in (\ref{4.10}) is gotten
theoretically in a natural way. We find that $j^{\mu _1\cdot \cdot \cdot \mu _k}\neq 0$ only when $\vec{\phi}%
=0$, which is just the singularity of $j^{\mu _1\cdot \cdot \cdot \mu _k}$.
In detail, the Kernel of the $\phi $--mapping is the singularities of the
topological tensor current $j^{\mu _1\cdot \cdot \cdot \mu _k}$ in $G$, i.e.
the inner structure of the topological tensor current is labelled by the zeroes
of $\phi$-mapping. We
think that this is the essential of the topological tensor current theory
and $\phi $--mapping is the key to study this theory.

From the above discussions, we see that the kernel of $\phi$--mapping plays an
important role in the topological tensor current theory, so we are focussed on
the zero points of $\vec \phi$ and will search for 
the solutions of the equations $\phi^a (x)=0$ $(a=1,...,m)$ by means of the 
implicit function theorem. These points are topological singularities in the
orientation of the order parameter field $\vec \phi (x)$. 

Suppose that the vector field $\vec{\phi}(x)$
possesses $l$ zeroes, according to the implicit function theorem\cite
{goursat1}, when the zeroes are regular points of $\phi $--mapping at which
the rank of the Jacobian matrix $[\partial _\mu \phi ^a]$ is $m$, the
solutions of $\vec{\phi}=0$ can be expressed parameterizedly by 
\begin{equation}
x^\mu =z_i^\mu (u^1,\cdot \cdot \cdot ,u^k),\;\;\;\;i=1,...,l,  \label{4.5}
\end{equation}
where the subscript $i$ represents the $i$--th solution and the parameters $%
u^I$ ($I=1,...,k$) span a $k$--dimensional submanifold with the metric
tensor $g_{IJ}=g_{\mu \nu }\frac{\partial x^\mu }{\partial u^I}\frac{%
\partial x^\nu }{\partial u^J}$ which is called the $i$--th singular
submanifold $N_i$ in the Riemannian manifold $G$ corresponding to the $\phi $%
--mapping. These singular submanifolds $N_i$ are just the world volumes
of the topological defects. For each singular manifold $N_i$, we can define a normal
submanifold $M_i$ in $G$ which is spanned by the parameters $v^A$ with the
metric tensor $g_{AB}=g_{\mu \nu }\frac{\partial x^\mu }{\partial v^A}\frac{%
\partial x^\nu }{\partial v^B}$ $(A,B=1,...,m)$, and the intersection point
of $M_i$ and $N_i$ is denoted by $p_i$ which can be expressed
parameterizedly by $v^A=p_i^A$. In fact, in the words of differential
topology, $M_i$ is transversal to $N_i$ at the point $p_i$. By virtue of the
implicit function theorem at the regular point $p_i$, it should be held true
that the Jacobian matrices $J(\frac \phi v)$ satisfies 
\begin{equation}
J(\frac \phi v)=\frac{D(\phi ^1,\cdot \cdot \cdot ,\phi ^m)}{D(v^1,\cdot
\cdot \cdot ,v^m)}\neq 0.  \label{4.nonzero}
\end{equation}

In the following, we will investigate the topological charges of the 
topological defects and their quantization. Let $\Sigma _i$ be a
neighborhood of $p_i$ on $M_i$ with boundary $\partial \Sigma _i$ 
satisfying $p_i\notin \partial \Sigma_i $, 
$\Sigma _i\cap \Sigma _j=\emptyset $. 
Then the generalized winding number $W_i$ of $n^a (x)$ at $p_i$\cite{dubrosin1}
can be defined by the Gauss map $n:\partial \Sigma _i\rightarrow S^{m-1}$
\begin{equation}
W_i=\frac 1{A(S^{m-1})(m-1)!}\int_{\partial \Sigma _i}n^{*}(\epsilon
_{a_1\cdot \cdot \cdot a_m}n^{a_1}dn^{a_2}\wedge \cdot \cdot \cdot \wedge
dn^{a_m})  \label{4.6}
\end{equation}
where $n^{*}$ denotes
the pull back of map $n$. The generalized 
winding numbers is a topological invariant and is also called the
degree of Gauss map\cite{milnor1}. It means that, when the point $v^A$ covers 
$\partial \Sigma
_i $ once, the unit vector $n^a$ will cover a region $n[\partial \Sigma _i]$
whose area is $W_i$ times of $A(S^{m-1})$, i.e. the unit vector $n^a$ will
cover the unit sphere $S^{m-1}$ for $W_i$ times. Using the Stokes'
theorem in exterior differential form and duplicating the above process,
we get the compact form of $W_i$
\begin{equation}
W_i=\int_{\Sigma _i}\delta (\vec{\phi})J(\frac \phi v)d^mv.  \label{4.W}
\end{equation}
By analogy with the procedure of deducing $\delta (f(x))$, since 
\begin{equation}
\delta (\vec{\phi})=\left\{ 
\begin{array}{cc}
+\infty , & for\;\vec{\phi}(x)=0 \\ 
0, & for\;\vec{\phi}(x)\neq 0
\end{array}
\right. =\left\{ 
\begin{array}{cc}
+\infty , & for\;x\in N_i \\ 
0, & for\;x\notin N_i
\end{array}
\right. ,
\end{equation}
we can expand the $\delta $--function $\delta (\vec{\phi})$ as 
\begin{equation}
\delta (\vec{\phi})=\sum_{i=1}^lc_i\delta (N_i),  \label{4.delta}
\end{equation}
where the coefficients $c_i$ must be positive, i.e. $c_i=\mid c_i\mid $.
$\delta (N_i)$ is the $\delta $--function in
space-time $G$ on a 
submanifold $N_i$\cite{gelfand1}
\begin{equation}
\delta (N_i)=\int_{N_i}\frac 1{\sqrt{g_x}}\delta ^n(\vec{x}-\vec{z}%
_i(u^1,\cdot \cdot \cdot, u^k))\sqrt{g_u}d^ku,  \label{4.m}
\end{equation}
where $g_x=\det (g_{\mu \nu })$, $g_u=\det (g_{IJ})$.
Substituting 
(\ref{4.delta}) into (\ref{4.W}), and calculating the integral,
we get the expression of $c_i$
\begin{equation}
c_i=\frac{\beta _i\sqrt{g_v}}{\mid J(\frac \phi v)_{p_i}\mid }=\frac{\beta
_i\eta _i\sqrt{g_v}}{J(\frac \phi v)_{p_i}},
\;\;\;g_v=det(g_{AB}),
\end{equation}
where $\beta _i=|W_i|$ is a positive integer called the Hopf index\cite
{milnor1} of $\phi $-mapping on $M_i,$ it means that when the point $v$
covers the neighborhood of the zero point $p_i$ once, the function $\vec{\phi%
}$ covers the corresponding region in $\vec{\phi}$-space $\beta _i$ times,
and $\eta _i=signJ(\frac \phi v)_{p_i}=\pm 1$ is the Brouwer degree of $\phi 
$-mapping\cite{milnor1}. Substituting this expression of $c_i$ and (\ref
{4.delta}) into (\ref{4.10}), we gain the total expansion of the rank--$k$
topological current 
\[
j^{\mu _1\cdot \cdot \cdot \mu _k}=\frac 1{\sqrt{g_x}}\sum_{i=1}^l\frac{%
\beta _i\eta _i\sqrt{g_v}}{J(\frac \phi v)|_{p_i}}\delta (N_i)J^{\mu _1\cdot
\cdot \cdot \mu _k}(\frac \phi x). 
\]
or in terms of parameters $y^{A^{^{\prime }}}=(v^1,\cdot \cdot \cdot
,v^m,u^1,\cdot \cdot \cdot ,u^k)$%
\begin{equation}
j^{A_1^{^{\prime }}\cdot \cdot \cdot A_k^{^{\prime }}}=\frac 1{\sqrt{g_y}}%
\sum_{i=1}^l\frac{\beta _i\eta _i\sqrt{g_v}}{J(\frac \phi v)|_{p_i}}\delta
(N_i)J^{A_1^{^{\prime }}\cdot \cdot \cdot A_k^{^{\prime }}}(\frac \phi x).
\end{equation}
From the above equation, we conclude that the inner structure of $j^{\mu
_1\cdot \cdot \cdot \mu _k}$ or $j^{A_1^{^{\prime }}\cdot \cdot \cdot
A_k^{^{\prime }}}$ is labelled by the total expansion of $\delta (\vec{\phi}%
) $, which includes the topological information $\beta _i$ and $\eta _i.$

Taking $u^1$ and $u^I$ $(I=2,...,k)$ be time-like evolution parameter and
space-like parameters,
respectively, the inner structure of the generalized topological current
just represents $l$ 
$(k-1)$--dimensional topological defects with topological charges
$g_i=\beta _i \eta _i$ moving in the $n$--dimensional
Riemann manifold $G$. The $k$-dimensional singular submanifolds $%
N_i\,\,(i=1,\cdot \cdot \cdot l)$ are their world sheets in the space-time.
Mazenko\cite{nuno} and Halperin\cite{halperin} also got similar results
for the case of point-like defects and line defects, but unfortunately, they
did not consider the case $\beta _i\neq 1$. In fact, what they lost sight
of is just the most important topological information for the charge of
topological defecs. In detail,
the Hopf indices $\beta _i$ characterize the absolute values of the
topological charges of these defects and the Brouwer degrees $\eta _i=+1$ correspond to
defects while $\eta _i=-1$ to antidefects. Furthermore, they did not discuss
what will happen when $\eta _i$ is indefinite, which we will study in detail
in section 3.

Corresponding to the rank--$k$ topological tensor currents $j^{\mu _1\cdot
\cdot \cdot \mu _k}$, it is easy to see that the Lagrangian of many defects
is just 
\[
L=\sqrt{\frac 1{k!}g_{\mu _1\nu _1}\cdot \cdot \cdot g_{\mu _k\nu _k}j^{\mu
_1\cdot \cdot \cdot \mu _k}j^{\nu _1\cdot \cdot \cdot \nu _k}}=\delta (\vec{%
\phi}) 
\]
which includes the total information of arbitrary dimensional topological
defects in $G$ and is the generalization of Nielsen's Lagrangian\cite
{nielsen1}. The action in $G$ is expressed by 
\[
S=\int_GL\sqrt{g_x}d^nx=\sum_{i=1}^l\beta _i\eta _i\int_{N_i}\sqrt{g_u}%
d^ku=\sum_{i=1}^l\beta _i\eta _iS_i 
\]
where $S_i$ is the area of the singular manifold $N_i$. It must be pointed
out here that the Nambu--Goto action\cite{nambu1}, which is the basis of
many works on defect theory, is derived naturally from our theory. From the
principle of least action, we obtain the evolution equations of many defect
objects 
\begin{equation}
\frac 1{\sqrt{g_u}}\frac \partial {\partial u^I}(\sqrt{g_u}g^{IJ}\frac{%
\partial x^\nu }{\partial u^J})+g^{IJ}\Gamma _{\mu \lambda }^\nu \frac{%
\partial x^\mu }{\partial u^I}\frac{\partial x^\lambda }{\partial u^J}%
=0,\;\;\;I,J=1,...,k.  \label{4.38}
\end{equation}
As a matter of fact, this is just the equation of harmonic map\cite{duan7}.

\section{The branch processes of the topological defects}

With the discussion mentioned above, we know that the results in the above
section are
obtained straightly from the topological view point under the condition $J(\phi/v)|_{p_i}\neq 0$, i.e. at the regular
points of the order parameter field $\vec \phi$. When the condition fails,
i.e. the Brouwer degree $\eta_i$ are indefinite, what will happen? In what
follows, we will study the case when $J(\phi/v)|_{p_i}=0$. It often happens
when the zero points of field $\vec \phi$ include some branch points, which
lead to the bifurcation of the topological current.

In this section, we will discuss the branch processes of these topological
defects. In order to simplify our study, we select the parameter $u^1$ as
the time--like evolution parameter $t$, and let the space--like parameters $%
u^I=\sigma ^I\;(I=2,...,k)$ be fixed. In this case, the Jacobian matrices $%
J^{A_1^{^{\prime }}\cdot \cdot \cdot A_k^{^{\prime }}}(\frac \phi y)$ are
reduced to 
\[
J^{AI_1\cdot \cdot \cdot I_{k-1}}(\frac \phi y)\equiv J^A(\frac \phi y%
),\;\;\;\;J^{ABI_1\cdot \cdot \cdot I_{k-2}}(\frac \phi y)=0,\;\;\;%
\;J^{(m+1)\cdot \cdot \cdot n}(\frac \phi y)=J(\frac \phi v), 
\]
\begin{equation}
A,B=1,...,(m+1),\;\;\;\;I_j=m+2,...,n,
\end{equation}
for $y^A=v^A\;(A\leq m),\;y^{m+1}=t,\;y^{m+I}=\sigma ^I\;(I\geq 2)$. The
branch points are determined by the $m+1$ equations 
\begin{equation}
\phi ^a(v^1,\cdots ,v^m,t,\vec{\sigma})=0,\;\;\;a=1,...,m  \label{4.phia}
\end{equation}
and 
\begin{equation}
\phi ^{m+1}(v^1,\cdots ,v^m,t,\vec{\sigma})\equiv J(\frac \phi v)=0
\label{4.zero}
\end{equation}
for the fixed $\vec{\sigma}$. and they are denoted as $(t^{*},p_i)$. In $%
\phi $--mapping theory usually there are two kinds of branch points, namely
the limit points and bifurcation points\cite{kubicek1}, satisfying 
\begin{equation}
J^1(\frac \phi y)|_{(t^{*},p_i)}\neq 0  \label{4.nonzero1}
\end{equation}
and 
\begin{equation}
J^1(\frac \phi y)|_{(t^{*},p_i)}=0,  \label{4.zero1}
\end{equation}
respectively. In the following, we assume that the branch points $%
(t^{*},p_i) $ of $\phi $--mapping have been found.

\subsection{Branch process at the limit point}

We first discuss the branch process at the limit point satisfying the 
condition (\ref{4.nonzero1}). In order to use the theorem of implicit function to study the branch process
of topological defects at the limit point, we use the Jacobian $J^1(\frac %
\phi y)$ instead of $J(\frac \phi v)$ to discuss the problem. In fact, this
means that we have replaced the parameter $t$ by $v^1$. For clarity we
rewrite the problem as 
\begin{equation}
\phi ^a(t,v^2,\cdots ,v^m,v^1,\vec{\sigma})=0,\;\;\;\;\;a=1,...,m.
\label{4.101}
\end{equation}
Then, taking account of the condition(\ref{4.nonzero1}) and using the
implicit function theorem, we have an unique solution of the equations (\ref
{4.101}) in the neighborhood of the limit point $(t^{*},p_i)$%
\begin{equation}
t=t(v^1,\vec{\sigma}),\;\;\;\;v^i=v^i(v^1,\vec{\sigma}),\;\;\;\;i=2,3,...,m
\label{4.102}
\end{equation}
with $t^{*}=t(p_i^1,\vec{\sigma})$. In order to show the behavior of the
defects at the limit points, we will investigate the Taylor expansion of (%
\ref{4.102}) in the neighborhood of $(t^{*},p_i)$. In the present case, from
(\ref{4.nonzero1}) and (\ref{4.zero}), we get 
\[
\frac{dv^1}{dt}|_{(t^{*},p_i)}=\frac{J^1(\frac \phi y)}{J(\frac \phi y)}%
|_{(t^{*},p_i)}=\infty , 
\]
i.e. 
\[
\frac{dt}{dv^1}|_{(t^{*},p_i)}=0. 
\]
Then we have the Taylor expansion of (\ref{4.102}) at the point $(t^{*},p_i)$%
\[
t=t(p_i,\vec{\sigma})+\frac{dt}{dv^1}|_{(t^{*},p_i)}(v^1-p_i^1)+\frac 12%
\frac{d^2t}{(dv^1)^2}|_{(t^{*},p_i)}(x^1-p_i^1)^2 
\]
\[
=t^{*}+\frac 12\frac{d^2t}{(dv^1)^2}|_{(t^{*},p_i)}(v^1-p_i^1)^2. 
\]
Therefore 
\begin{equation}
t-t^{*}=\frac 12\frac{d^2t}{(dv^1)^2}|_{(t^{*},p_i)}(v^1-p_i^1)^2
\label{4.103}
\end{equation}
which is a parabola in the $v^1$---$t$ plane. From (\ref{4.103}), we can
obtain the two solutions $v_{(1)}^1(t,\vec{\sigma})$ and $v_{(2)}^1(t,\vec{%
\sigma})$, which give the branch solutions of the system (\ref{4.phia}) at
the limit point. If $\frac{d^2t}{(dv^1)^2}|_{(t^{*},p_i)}>0$, we have the
branch solutions for $t>t^{*}$ (Fig 1(a)), otherwise, we have the branch solutions for $%
t<t^{*} $ (Fig 1(b)). The former is related to the
creation of defect and antidefect in pair at the limit points, and the latter
to the annihilation of the topological defects, since the topological current of topological defects
is identically conserved, the topological quantum numbers of these
two generated topological defects must be opposite at the limit point,
i.e. $\beta _1\eta _1+\beta _2\eta _2=0$.

\subsection{Branch process at the bifurcation point}

In the following, let us consider the case (\ref{4.zero1}), in which the
restrictions of the system (\ref{4.phia}) at the bifurcation point $%
(t^{*},p_i) $ are 
\begin{equation}  \label{4.104}
J(\frac \phi v)|_{(t^{*},p_i)}=0,\;\;\;J(\frac \phi v)|_{(t^{*},p_i)}=0.
\end{equation}
These two restrictive conditions will lead to an important fact that the
dependency relationship between $t$ and $v^1$ is not unique in the
neighborhood of the bifurcation point $(t^{*},p_i).$ In fact, we have 
\begin{equation}  \label{4.105}
\frac{dv^1}{dt}|_{(t^{*},p_i)}=\frac{J^1(\frac \phi y)}{J(\frac \phi v)}%
|_{(t^{*},p_i)}
\end{equation}
which under the restraint (\ref{4.104}) directly shows that the tangential
direction of the integral curve of equation (\ref{4.105}) is indefinite at
the point $(t^{*},p_i)$. Hence, (\ref{4.105}) does not satisfy the
conditions of the existence and uniqueness theorem of the solution of a
differential equation. This is why the very point $(t^{*},\vec z_i)$ is
called the bifurcation point of the system (\ref{4.phia}).

In the following, we will find a simple way to search for the different
directions of all branch curves at the bifurcation point. As assumed that
the bifurcation point $(t^{*},p_i)$ has been found from (\ref{4.phia}) and (%
\ref{4.zero}), the following calculations are all conducted at the value $%
(t^{*},p_i)$. As we have mentioned above, at the bifurcation point $%
(t^{*},p_i)$, the rank of the Jacobian matrix $[\frac{\partial \phi }{%
\partial v}]$ is smaller than $m$. In order to derive the calculating
method, we consider the rank of the Jacobian matrix $[\frac{\partial \phi }{%
\partial v}]$ is $m-1$. The case of a more smaller rank will be discussed in
next subsection. Suppose that one of the $(m-1)\times (m-1)$ submatrix $J_1(%
\frac \phi v)$ of the Jacobian matrix $[\frac{\partial \phi }{\partial v}]$
is 
\begin{equation}  \label{4.106}
J_1(\frac \phi v)=\left( 
\begin{array}{cccc}
\phi _2^1 & \phi _3^1 & \cdots & \phi _m^1 \\ 
\phi _2^2 & \phi _3^2 & \cdots & \phi _m^2 \\ 
\vdots & \vdots & \ddots & \vdots \\ 
\phi _2^{m-1} & \phi _3^{m-1} & \cdots & \phi _m^{m-1}
\end{array}
\right)
\end{equation}
and its determinant $\det J_1(\frac \phi v)$ does not vanish at the point $%
(t^{*},p_i)$ (otherwise, we have to rearrange the equations of (\ref{4.phia}%
)), where $\phi _A^a$ stands for $(\partial \phi ^a/\partial v^A)$ $%
(a=1,...,m-1;\;A=2,...,m)$. By means of the implicit function theorem we
obtain one and only one functional relationship in the neighborhood of the
bifurcation point $(t^{*},p_i)$%
\begin{equation}  \label{4.107}
v^A=f^A(v^1,t,\sigma ^2,\cdots ,\sigma ^k),\;\;\;\;\;A=2,3,...,n
\end{equation}
with the partial derivatives 
\[
f_1^A=\frac{\partial v^A}{\partial v^1},\;\;\;f_t^A=\frac{\partial v^A}{%
\partial t},\;\;\;A=2,3,...,n. 
\]
Then, for $a=1,...,m-1$ we have 
\[
\phi ^a=\phi ^a(v^1,f^2(v^1,t,\vec \sigma ),...,f^m(v^1,t,\vec \sigma ),t,%
\vec \sigma )\equiv 0 
\]
which gives 
\begin{equation}  \label{4.108}
\sum\limits_{A=2}^m\frac{\partial \phi ^a}{\partial v^A}f_1^A=-\frac{%
\partial \phi ^a}{\partial v^1},\;\;\;a=1,...,m-1
\end{equation}
\begin{equation}  \label{4.109}
\sum\limits_{A=2}^m\frac{\partial \phi ^a}{\partial v^A}f_t^A=-\frac{%
\partial \phi ^a}{\partial t},\;\;\;a=1,...,m-1
\end{equation}
from which we can calculate the first order derivatives of $f^A$ : $f_1^A$
and $f_t^A$. Denoting the second order partial derivatives as 
\[
f_{11}^A=\frac{\partial ^2v^A}{(\partial v^1)^2},\;\;f_{1t}^A=\frac{\partial
^2v^A}{\partial v^1\partial t},\;\;\;f_{tt}^A=\frac{\partial ^2v^A}{\partial
t^2} 
\]
and differentiating (\ref{4.108}) with respect to $v^1$ and $t$
respectively, we get 
\begin{equation}  \label{4.110}
\sum\limits_{A=2}^m\phi _A^af_{11}^A=-\sum\limits_{A=2}^m[2\phi
_{A1}^af_1^A+\sum\limits_{B=2}^m(\phi _{AB}^af_1^B)f_1^A]-\phi
_{11}^a,\;\;\;a=1,2,...,m-1
\end{equation}
\begin{equation}  \label{4.111}
\sum\limits_{A=2}^m\phi _A^af_{1t}^A=-\sum\limits_{A=2}^m[\phi
_{At}^af_1^A+\phi _{A1}^af_t^A+\sum\limits_{B=2}^m(\phi
_{AB}^af_t^B)f_1^A]-\phi _{1t}^a,\;\;\;a=1,2,...,m-1.
\end{equation}
And the differentiation of (\ref{4.109}) with respect to $t$ gives 
\begin{equation}  \label{4.112}
\sum\limits_{A=2}^m\phi _A^af_{tt}^A=-\sum\limits_{A=2}^m[2\phi
_{At}^af_t^A+\sum\limits_{B=2}^m(\phi _{AB}^af_t^B)f_t^A]-\phi
_{tt}^a,\;\;\;a=1,2,...,m-1
\end{equation}
where 
\[
\phi _{AB}^a=\frac{\partial ^2\phi ^a}{\partial v^A\partial v^B},\;\;\;\phi
_{At}^a=\frac{\partial ^2\phi ^a}{\partial v^A\partial t}. 
\]
The differentiation of (\ref{4.109}) with respect to $v^1$ gives the same
expression as (\ref{4.111}). If we use the Gaussian elimination method to
the three vectors at the right hands of the formulas (\ref{4.110}), (\ref
{4.111}) and (\ref{4.112}), we can obtain the three partial derivatives $%
f_{11}^A,\;f_{1t}^A$ and $f_{tt}^A$. Notice that the three equations (\ref
{4.110}), (\ref{4.111}) and (\ref{4.112}) have the same coefficient matrix $%
J_1(\frac \phi v)$, which are assumed to be nonzero, and we should
substitute the values of the partial derivatives $f_1^A$ and $f_t^A$, which
have been calculated out in the former, into the right hands of the three
equations.

The above discussions do not matter to the last component $\phi ^m(v^1,\cdot
\cdot \cdot ,v^m,t,\vec \sigma ).\,$ In order to find the different values
of $dv^1/dt$ at the bifurcation point, let us investigate the Taylor
expansion of $\phi ^m(v^1,\cdot \cdot \cdot ,v^m,t,\vec \sigma )$ in the
neighborhood of $(t^{*},p_i)$. Substituting the existing, but unknown,
dependency relationship (\ref{4.107}) into $\phi ^m(v^1,\cdot \cdot \cdot
,v^m,t,\vec \sigma )$, we get the function of two variables $v^1$ and $t$%
\begin{equation}  \label{4.113}
F(t,v^1,\vec \sigma )=\phi ^m(v^1,f^2(v^1,t,\vec \sigma ),...,f^m(v^1,t,\vec %
\sigma ),t,\vec \sigma )
\end{equation}
which according to (\ref{4.phia}) must vanish at the bifurcation point 
\begin{equation}  \label{4.114}
F(t^{*},p_i)=0.
\end{equation}
From (\ref{4.113}), we can calculate the first order partial derivatives of $%
F(t,v^1,\vec \sigma )$ with respect to $v^1$ and $t$ respectively at the
bifurcation point $(t^{*},p_i)$%
\begin{equation}  \label{4.115}
\frac{\partial F}{\partial v^1}=\phi _1^m+\sum\limits_{A=2}^m\phi
_A^mf_1^A,\;\;\;\frac{\partial F}{\partial t}=\phi
_t^m+\sum\limits_{A=2}^m\phi _A^mf_t^A.
\end{equation}
Using (\ref{4.108}) and (\ref{4.109}), the first equation of (\ref{4.104})
is expressed by 
\[
J(\frac \phi v)|_{(t^{*},p_i)}=\left| 
\begin{array}{cccc}
-\sum\limits_{A=2}^m\phi _A^1f_1^A & \phi _2^1 & \cdots & \phi _m^1 \\ 
-\sum\limits_{A=2}^m\phi _A^2f_1^A & \phi _2^2 & \cdots & \phi _m^2 \\ 
\vdots & \vdots & \cdot & \vdots \\ 
-\sum\limits_{A=2}^m\phi _A^{m-1}f_1^A & \phi _2^{m-1} & \cdots & \phi
_m^{m-1} \\ 
\phi _A^m & \phi _2^m & \cdots & \phi _m^m
\end{array}
\right| _{(t^{*},p_i)}=0 
\]
which, by Cramer's rule, (\ref{4.106}) and (\ref{4.115}), can be rewritten
as 
\[
J(\frac \phi v)|_{(t^{*},p_i)}=\left| 
\begin{array}{cccc}
0 & \phi _2^1 & \cdots & \phi _m^1 \\ 
0 & \phi _2^2 & \cdots & \phi _m^2 \\ 
\vdots & \vdots & \ddots & \vdots \\ 
0 & \phi _2^{m-1} & \cdots & \phi _m^{m-1} \\ 
\phi _1^m+\sum\limits_{A=2}^m\phi _A^mf_1^A & \phi _2^m & \cdots & \phi _m^m
\end{array}
\right| _{(t^{*},p_i)} 
\]
\[
=\frac{\partial F}{\partial v^1}\det J_1(\frac \phi v)|_{(t^{*},p_i)}=0. 
\]
Since 
\[
\det J_1(\frac \phi v)|_{(t^{*},p_i)}\neq 0 
\]
which is our assumption, the above equation leads to 
\begin{equation}  \label{4.116}
\frac{\partial F}{\partial v^1}|_{(t^{*},p_i)}=0.
\end{equation}
With the same reasons, we can prove that 
\begin{equation}  \label{4.117}
\frac{\partial F}{\partial t}|_{(t^{*},p_i)}=0.
\end{equation}
The second order partial derivatives of the function $F(t,v^1,\vec \sigma )$
are easily to find out to be 
\[
\frac{\partial ^2F}{(\partial v^1)^2}=\phi _{11}^m+\sum\limits_{A=2}^m[2\phi
_{1A}^mf_1^A+\phi _A^mf_{11}^A+\sum\limits_{B=2}^m(\phi _{AB}^mf_1^B)f_1^A] 
\]
\[
\frac{\partial ^2F}{\partial v^1\partial t}=\phi
_{1t}^m+\sum\limits_{A=2}^m[\phi _{1A}^mf_t^A+\phi _{tA}^mf_1^A+\phi
_A^mf_{1t}^A+\sum\limits_{B=2}^m(\phi _{AB}^mf_t^B)f_1^A] 
\]
\[
\frac{\partial ^2F}{\partial t^2}=\phi _{tt}^m+\sum\limits_{A=2}^m[2\phi
_{At}^mf_t^A+\phi _A^mf_{tt}^A+\sum\limits_{B=2}^m(\phi _{AB}^mf_t^B)f_t^A] 
\]
which at $(t^{*},p_i)$ are denoted by 
\begin{equation}  \label{4.118}
A=\frac{\partial ^2F}{(\partial v^1)^2}\mid _{(t^{*},p_i)},\quad \quad B=%
\frac{\partial ^2F}{\partial v^1\partial t}\mid _{(t^{*},p_i)},\ \quad \quad
C=\frac{\partial ^2F}{\partial t^2}\mid _{(t^{*},p_i)}.
\end{equation}
Then, by virtue of (\ref{4.114}), (\ref{4.116}), (\ref{4.117}) and (\ref
{4.118}), the Taylor expansion of $F(t,v^1,\vec \sigma )$ in the
neighborhood of the bifurcation point $(t^{*},p_i)$ can be expressed as 
\begin{equation}  \label{4.119}
F(t,v^1,\vec \sigma )=\frac 12A(v^1-p_i^1)^2+B(v^1-p_i^1)(t-t^{*})+\frac 12%
C(t-t^{*})^2
\end{equation}
which is the expression of $\phi ^m(v^1,\cdots ,v^m,t,\vec \sigma )$ in the
neighborhood of $(t^{*},p_i)$. The expression (\ref{4.119}) shows that at
the bifurcation point $(t^{*},p_i)$ 
\begin{equation}  \label{4.120}
A(v^1-p_i^1)^2+2B(v^1-p_i^1)(t-t^{*})+C(t-t^{*})^2=0.
\end{equation}
Dividing (\ref{4.120}) by $(v^1-p_i^1)^2$ or $(t-t^{*})^2$, and taking the
limit $t\rightarrow t^{*}$ as well as $v^1\rightarrow p_i^1$ respectively,
we get two equations 
\begin{equation}  \label{4.121}
C(\frac{dt}{dv^1})^2+2B\frac{dt}{dv^1}+A=0.
\end{equation}
and 
\begin{equation}  \label{4.122}
A(\frac{dv^1}{dt})^2+2B\frac{dv^1}{dt}+C=0.
\end{equation}
So we get the different directions of the branch curves at the bifurcation
point from the solutions of (\ref{4.121}) or (\ref{4.122}). There are four
possible cases:

Firstly, $A \neq 0,\,$ $\Delta =4(B ^2-AC )>0$, from Eq. (\ref{4.121}) we
get two different solutions: $dv^1/dt\mid _{1,2}=(-B \pm \sqrt{B ^2-AC })/A $%
, which is shown in Fig. 2, where two topological defects meet and then depart at the
bifurcation point. Secondly, $A \neq 0,\,\Delta =4(B ^2-A C )=0$, there is
only one solution: $dv^1/dt=-B /A $, which includes three important cases:
(a) two topological defects tangentially collide at the bifurcation point (Fig 3(a));
(b) two topological defects merge into one topological defect at the bifurcation
point (Fig 3(b)); (c) one topological defect splits into two topological defects at
the bifurcation point (Fig 3(c)). Thirdly, $A =0,\,C \neq 0,$ $\Delta =4(B ^2-A C )>0$,
from Eq. (\ref{4.122}) we have $dt/dv^1=0$ and $-2B /C $. There are two
important cases: (i) One topological defect splits into three topological
defects at the bifurcation point (Fig 4(a)); (ii) Three topological defects merge into
one at the bifurcation point (Fig 4(b)). Finally, $A =C =0$, Eqs. (\ref{4.121}) and (%
\ref{4.122}) give respectively $dv^1/dt=0$ and $dt/dv^1=0$. This case is obvious as 
in Fig. 5, which is 
similar to the third situation.

In order to determine the branches directions of the remainder variables, we
will use the relations simply 
\[
dv^A=f_1^Adv^1+f_t^Adt,\;\;\;\;\;A=2,3,...,n 
\]
where the partial derivative coefficients $f_1^A$ and $f_t^A$ have given in (%
\ref{4.108}) and (\ref{4.109}). Then, respectively 
\[
\frac{dv^A}{dv^1}=f_1^A+f_t^A\frac{dt}{dv^1} 
\]
or 
\begin{equation}  \label{4.126}
\frac{dv^A}{dt}=f_1^A\frac{dv^1}{dt}+f_t^A.
\end{equation}
where partial derivative coefficients $f_1^A$ and $f_t^A$ are given by (\ref
{4.108}) and (\ref{4.109}). From this relations we find that the values of $%
dv^A/dt$ at the bifurcation point $(t^{*},z_i)$ are also possibly different
because (\ref{4.122}) may give different values of $dv^1/dt$.

\subsection{Branch process at a higher degenerated point}

In the following, let us discuss the branch process at a higher degenerated
point. In the above subsection, we have analysised the case that the rank of
the Jacobian matrix $[\partial \phi /\partial v]$ of the equation (\ref
{4.zero}) is $m-1$. In this section, we consider the case that the rank of
the Jacobian matrix is $m-2$ (for the case that the rank of the matrix $%
[\partial \phi /\partial v]$ is lower than $m-2$, the discussion is in the
same way). Let the $(m-2)\times (m-2)$ submatrix $J_2(\frac \phi v)$ of the
Jacobian matrix $[\partial \phi /\partial v]$ be 
\[
J_2(\frac \phi v)=\left( 
\begin{array}{cccc}
\phi _3^1 & \phi _4^1 & \cdots & \phi _m^1 \\ 
\phi _3^2 & \phi _4^2 & \cdots & \phi _m^2 \\ 
\vdots & \vdots & \ddots & \vdots \\ 
\phi _3^{m-2} & \phi _4^{m-2} & \cdots & \phi _m^{m-2}
\end{array}
\right) 
\]
and suppose that $\det J_2(\frac \phi v)|_{(t^{*},p_i)}\neq 0.$ With the
same reasons of obtaining (\ref{4.107}), we can have the function relations 
\begin{equation}  \label{4.127}
v^A=f^A(v^1,v^2,t,\vec \sigma ),\;\;\;\;\;A=3,4,...,m.
\end{equation}
For the partial derivatives $f_1^A$, $f_2^A$ and $f_t^A$, we can easily
derive the system similar to the equations (\ref{4.108}) and (\ref{4.109}),
in which the three terms at the right hand of can be figured out at the same
time. In order to determine the 2--order partial derivatives $f_{11}^A$, $%
f_{12}^A$, $f_{1t}^A$, $f_{22}^A$, $f_{2t}^A$ and $f_{tt}^A$, we can use the
equations similar to (\ref{4.110}), (\ref{4.111}) and (\ref{4.112}).
Substituting the relations (\ref{4.127}) into the last two equations of the
system (\ref{4.phia}), we have the following two equations with respect to
the arguments $v^1,\,\,v^2,\,\,t,\vec \sigma $%
\begin{equation}  \label{4.bifa46}
\left\{ 
\begin{array}{l}
F_1(v^1,v^2,t, \vec \sigma )=\phi ^{m-1}(v^1,v^2,f^3(v^1,v^2,t,\vec \sigma
),\cdots ,f^m(v^1,v^2,t,\vec \sigma ),t,\vec \sigma )=0 \\ 
F_2(v^1,v^2,t,\vec \sigma )=\phi ^m(v^1,v^2,f^3(v^1,v^2,t,\vec \sigma
),\cdots ,f^m(v^1,v^2,t,\vec \sigma ),t,\vec \sigma )=0.
\end{array}
\right.
\end{equation}
Calculating the partial derivatives of the function $F_1$ and $F_2$ with
respect to $v^1$, $v^2$ and $t$, taking notice of (\ref{4.127}) and using
six similar expressions to (\ref{4.116}) and (\ref{4.117}), i.e. 
\begin{equation}  \label{4.bifa48}
\frac{\partial F_j}{\partial v^1}\mid _{(t^{*},p_i)}=0,\ \quad \quad \frac{%
\partial F_j}{\partial v^2}\mid _{(t^{*},p_i)}=0,\ \quad \quad \frac{%
\partial F_j}{\partial t}\mid _{(t^{*},p_i)}=0,\ \quad \quad j=1,2,
\end{equation}
we have the following forms of Taylor expressions of $F_1$ and $F_2$ in the
neighborhood of $(t^{*},p_i)$ 
\[
F_j(v^1,v^2,t,\vec \sigma )\approx
A_{j1}(v^1-p_i^1)^2+A_{j2}(v^1-p_i^1)(v^2-p_i^2)+A_{j3}(v^1-p_i^1) 
\]
\[
(t-t^{*})+A_{j4}(v^2-p_i^2)^2+A_{j5}(v^2-p_i^2)(t-t^{*})+A_{j6}(t-t^{*})^2=0 
\]
\begin{equation}  \label{4.bifa49}
j=1,2.
\end{equation}
In the case of $A_{j1}\neq 0,A_{j4}\neq 0$, by dividing (\ref{4.bifa49}) by $%
(t-t^{*})^2$ and taking the limit $t\rightarrow t^{*}$, we obtain two
quadratic equations of $\frac{dv^1}{dt}$ and $\frac{dv^2}{dt}$ 
\begin{equation}  \label{4.bifa50}
A_{j1}(\frac{dv^1}{dt})^2+A_{j2}\frac{dv^1}{dt}\frac{dv^2}{dt}+A_{j3}\frac{%
dv^1}{dt}+A_{j4}(\frac{dv^2}{dt})^2+A_{j5}\frac{dv^2}{dt}+A_{j6}=0
\end{equation}
\[
j=1,2. 
\]
Eliminating the variable $dv^1/dt$, we obtain a equation of $dv^2/dt$ in the
form of a determinant 
\begin{equation}  \label{4.bifa51}
\left| 
\begin{array}{cccc}
A_{11} & A_{12}Q+A_{23} & A_{14}Q^2+A_{15}Q+A_{16} & 0 \\ 
0 & A_{11} & A_{12}Q+A_{13} & A_{14}Q^2+A_{15}Q+A_{16} \\ 
A_{21} & A_{22}Q+A_{23} & A_{24}Q^2+A_{25}Q+A_{26} & 0 \\ 
0 & A_{21} & A_{22}Q+A_{23} & A_{24}Q^2+A_{25}Q+A_{26}
\end{array}
\right| =0
\end{equation}
where $Q=dv^2/dt$, which is a $4th$ order equation of $dv^2/dt$ 
\begin{equation}  \label{4.bifa52}
a_0(\frac{dv^2}{dt})^4+a_1(\frac{dv^2}{dt})^3+a_2(\frac{dv^2}{dt})^2+a_3(%
\frac{dv^2}{dt})+a_4=0.
\end{equation}
Therefore we get different directions at the bifurcation point corresponding
to different branch curves. The number of different branch curves is four at
most. If the degree of degeneracy of the matrix $[\frac{\partial \phi }{%
\partial v}]$ is more higher, i.e. the rank of the matrix $[\frac{\partial
\phi }{\partial v}]$ is more lower than the present $(m-2)$ case, the
procedure of deduction will be more complicate. In general supposing the
rank of the matrix $[\frac{\partial \phi }{\partial x}]$ be $(m-s)$, the
number of the possible different directions of the branch curves is $2^s$ at
most.

At the end of this section, we conclude that there exist crucial cases of
branch processes in our topological defect theory. This means that a
topological defect, at the bifurcation point, may split into several (for
instance $s$) topological defects along different branch curves with
different charges. Since the topological current is a conserved current, the
total quantum number of the spliting topological defects must precisely
equal to the topological charge of the original defect i.e. 
\[
\sum\limits_{j=1}^s\beta _{i_j}\eta _{i_j}=\beta _i\eta _i 
\]
for fixed $i$. This can be looked upon as the topological reason of the
defect splitting. Here we should point out that such splitting is a
stochastic process, the sole restriction of this process is just the
conservation of the topological charge of the topological defects during
this splitting process. Of course, the topological charge of each splitting
defects is an integer.

In summary, we have studied the topological property of the arbitrary
dimensional topological defects in general case by making use of the $\phi $%
--mapping topological current theory and the implicit function theorem. We
would like to point out that all the results in this paper are gained from
the viewpoint of topology without any particular models or hypothesis.

\section*{Figures' Captions}

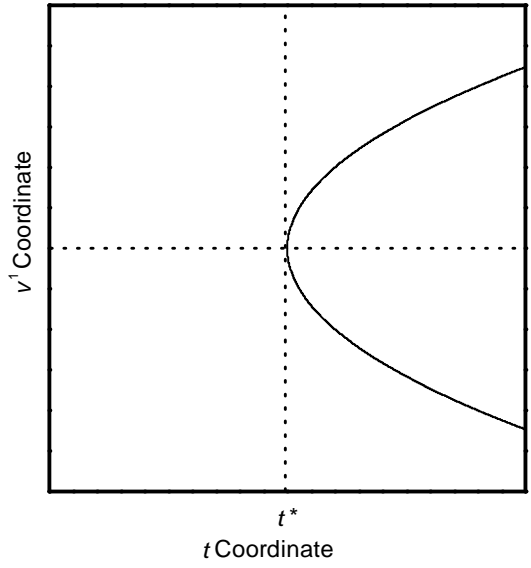
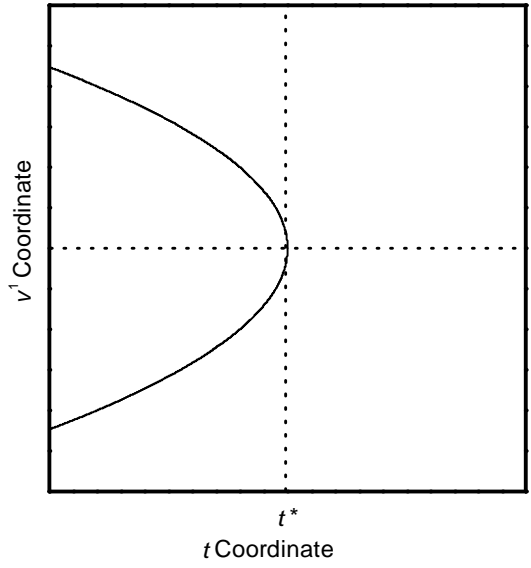
Fig. 1. (a) The creation of two topological defects. (b) Two 
topological defects
annihilate in collision at the limit point.

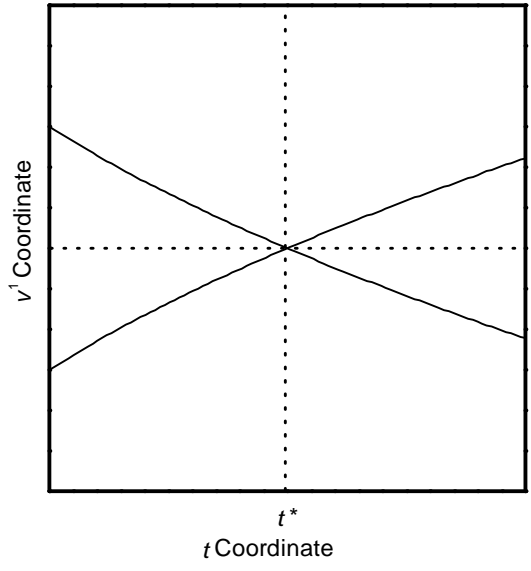
Fig. 2. Two topological defects collide with different directions of motion at
the bifurcation point.

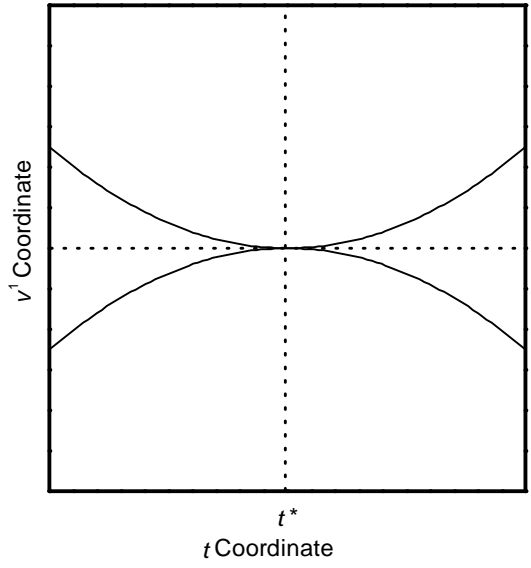
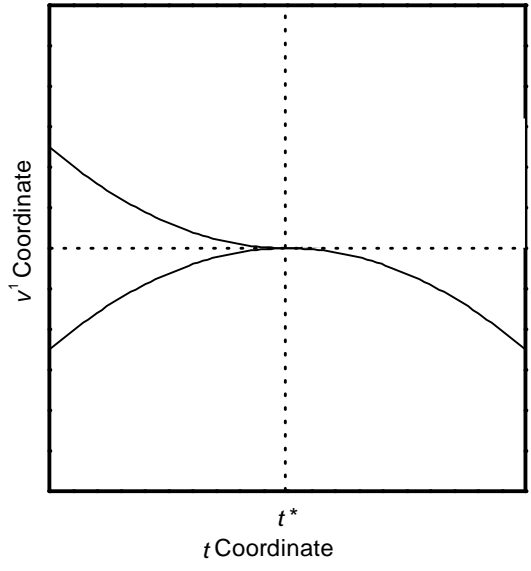
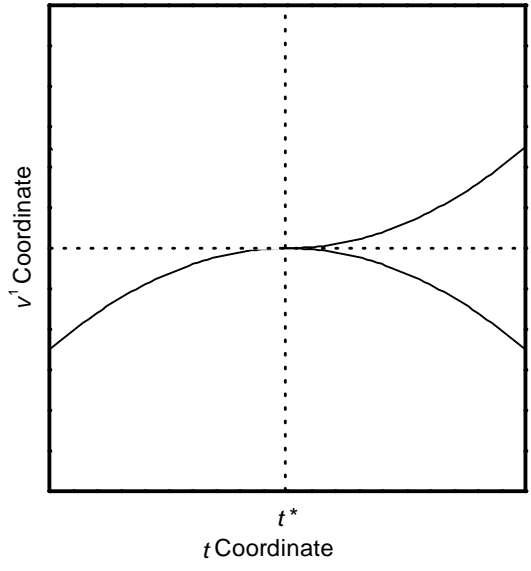
Fig. 3. Topological defects have the same direction of motion. (a) Two topological defects
tangentially collide at the bifurcation point. (b) Two topological defects
merge into one topological defect at the bifurcation point. (c)
One topological defect splits into two topological defects at the bifurcation
point.

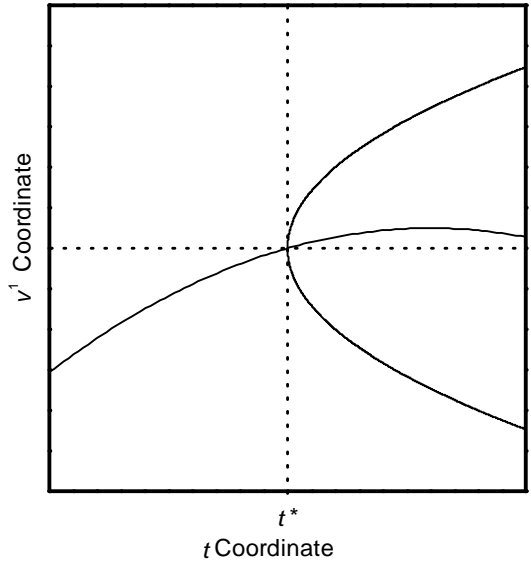
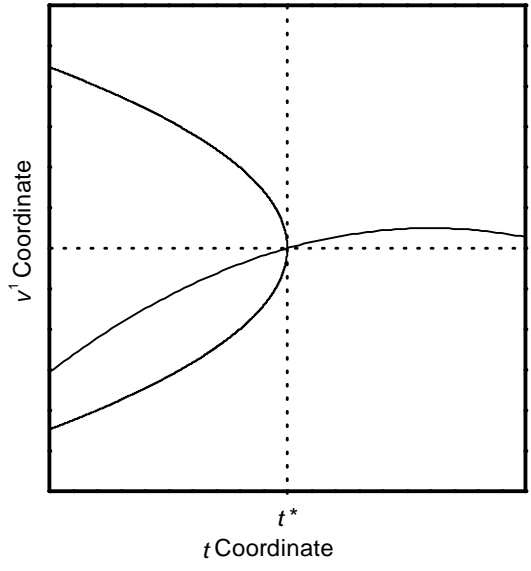
Fig. 4. (a) One topological defect splits into three topological defects at the
bifurcation point. (b) Three topological defects merge into one topological defect
at the bifurcation point.

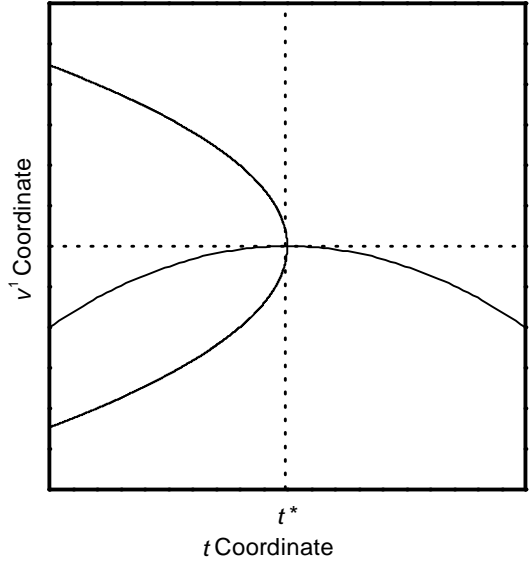
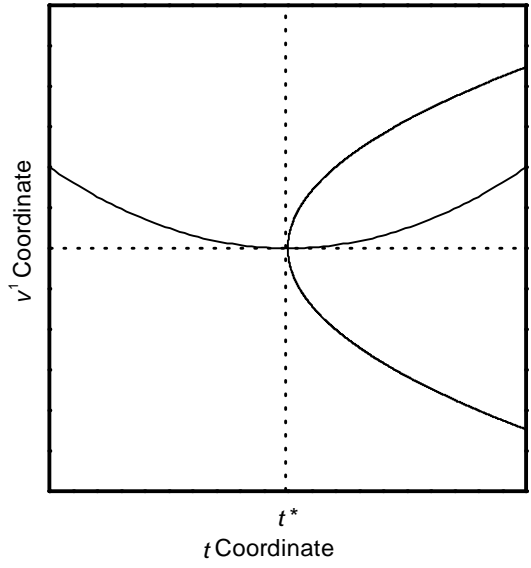
Fig. 5. This case is similar to Fig. 4. (a) Three topological defects merge
into one topological defect at the bifurcation point. (b) One topological defect
splits into three topological defects at the bifurcation point.

\end{document}